# Cluster-Impact Fusion and Effective Deuteron Temperature

**Y.E. Kim, J.-H. Yoon, and R. A. Rice**
*Department of Physics, Purdue University, West Lafayette, Indiana 47907*

**M. Rabinowitz**
*Electric Power Research Inst. and Armor Research, 715 Lakemead Way, Redwood City, CA 94062*

Inquiries to: Mario Rabinowitz; E-mail:  Mario715@earthlink.net

### Abstract

Temperature and kinematic line broadening are the  primary contributions to the width of the proton energy spectrum measured in cluster-impact fusion experiments.  By ascertaining these two contributions, we have determined an effective temperature for the high-velocity deuteron component that is responsible for the measured fusion yield.  The extracted effective temperature is substantially higher than conventional estimates., and implies that cluster-impact fusion is hot fusion on an atomic scale.  The proton spectrum rules out contaminants in explaining the high yield.

**PACS** numbers:  79.20.RF, 25.45.--z, 47.40.Nm, 52.50. Lp

Unexpected high fusion rates of ~ 1 - 10 $s^{-1}$/D-D observed in the recent cluster-impact fusion (CIF) experiments [1 - 3] met with considerable skepticism because of the expected low energies of deuterons in the projectile molecular clusters, $(D_2O)_n^+$.

The experimental yields for $(D_2O)_{100}^+$ clusters are ~ $10^{25}$ times higher than that expected

[4, 5] from single $D^+$ ions at D-D center-of-mass (c.m.) energies of 150 eV/D, and ~ $10^{100}$ times higher than at 15 eV/D for $(D_2O)_{1000}^+$. A number of theoretical models [6 -10] were proposed, but they greatly underestimated the observed fusion yields [1, 2, 4] for n > 100. It has been recently shown [8 - 10] that heavy atomic partners (such as the oxygen in $D_2O$) in the molecule play a vital role in explaining the apparently conflicting negative results for $(D)_n^+$ clusters [11]. The fusion yields due to light-fragment contaminants in the beam (e.g. $D^+$, $D_2O^+$, $D_3O^+$) are insufficient [12] to account for the experimental results given the determined upper limits [2] for these contaminants in the CIP experiments. Beuhler $et\ al.$ [13] have recently shown with their time-of-flight experiments that light-fragment contaminants cannot be responsible for observed fusion events.

Recently, it has been shown [12, 14 - 16] that all of the observed D -D fusion rates [2] for 150 - 300 keV $(D_2O)_{100}^+$ clusters striking TiD, $C_2D_4$, and $ZrD_{1.65}$ targets can be reproduced by using the effective translational (ET) temperature of $T_e$ (n = 100) = $E_t/46$ (3.3 - 6.5 keV for a cluster energy of $E_t$ = 150 - 300 keV) which is substantially larger than conventional estimates (< 700 eV) based on classical molecular-dynamics calculations using $(D_2O)_n$ [6, 17 - 19]. In this paper, we show that the ET temperature $T_e$ can be extracted directly from the experimentally measured proton energy spectrum as in the case of D -D fusion in a hot plasma for which the neutron energy spectrum has been used as a diagnostic of the temperture [20].

In CIF experiments, $(D_2O)_n^+$ molecular clusters impact on the deuterated target with a cluster velocity of ~ $10^7$ cm/sec in an impact period of $\Delta t \approx 10^{-14}$ sec. A one-dimensional (1D) velocity distribution yields a symmetric broadening consistent with the data [2]. We assume that at impact a fraction $f_D$ of projectile deuterons in the cluster $(D_2O)_n^+$ and target deuterons develop a 1 D distribution

$$f(v_z) = \frac{1}{\sqrt{\pi}} \left( \frac{m}{2kT_e} \right)^{1/2} e^{-mv_z^2/2kT_e} \qquad (1)$$

with an ET temperature or energy $E_e = kT_e$ (k is the Boltzmann constant). Note that

$$\int_{-\infty}^{\infty} f(v_z) dv_z = 1,$$

$$\overline{v}_z = \int_{-\infty}^{\infty} |v_z| f(v_z) dv_z = \left( 2kT_e / m\pi \right)^{1/2},$$

and $\quad \frac{1}{2} m \overline{v_z^2} = \int_{-\infty}^{\infty} \frac{1}{2} m v_z^2 f(v_z) dv_z = kT_e / 2.$

m is the deuteron rest mass. Although the total number of atoms is limited to 3n per

$(D_2O)_n^+$ cluster, the total number $N_t$ of atoms in a cluster beam with a current of ~ 1 nA is a large statistical system, $N_t = 0.625 \times 10^{10}$ (3n)/sec [$\approx 2 \times 10^{12}$ / sec for $(D_2O)_{100}^+$].

The total fusion proton yield per cluster due to the D(D,p)T reaction [12, 15] for the deuteron velocity distribution of Eq. (1) moving into the target after the cluster impact is given by

$$Y(n, E_t, kT_e) = 2ngf_D n_D' \int_0^{E_t} \frac{\sigma(E_{c.m.})}{|dE / dx|} \left( e^{-E/kT_e} - e^{-E_t/kT_e} \right) dE, \qquad (2)$$

where the target deuteron number density $n_D'$ is $5.68 \times 10^{22}$, $7.05 \times 10^{22}$, and $8 \times 10^{22}$ cm$^{-3}$ for the targets TiD, ZrD$_{1.65}$, and C$_2$D$_4$, respectively, and dE/dx is the stopping power of the target for a deuterium projectile [12, 21]. $\sigma(E_{c.m.})$ is the cross section for the D(D,p)T reaction with the c.m. deuteron kinetic energy $E_{c.m.} = E/2$. We use the conventional parametrization for the cross section [15, 22]. For the cases considered in this paper, $E_t$ in Eq. (2) is set to infinity without loss of accuracy. The factor g (set to 0.5) in Eq. (2) is included to account for the fraction of $f_D$ that is lost in the backward direction away from the target.

The proton spectrum measured by Beuhelr *et al.* [2] for the case of a $(D_2O)_{115}^+$ cluster with cluster energy $E_t = 275$ keV impacting the $C_2D_4$ target has a broad width; cf. Fig. 1. Several mechanisms contribute to line broadening of the proton energy spectrum. A dominant one is kinematic broadening due to the finite acceptance angle of the proton detector. For the observed full width at half maximum (FWHM) value of $\Delta E_p \approx 320$ keV [2], the circular proton detector had a surface area of 3 cm$^2$ and was placed 1 cm from the target perpendicular to the incident beam, for a detector orientation angle of $\Theta_D = 90^o$ and a maximum acceptance angle of $\Theta_L = 90^o \pm 44.34^o$, where $\Theta_L$ is the proton scattering angle with respect to the incident beam direction (the positive z axis) in the laboratory frame (i.e. $\Theta_L = 0^o$ for forward scattering). The deuterated targets were oriented at $\Theta_L = 45^o$.

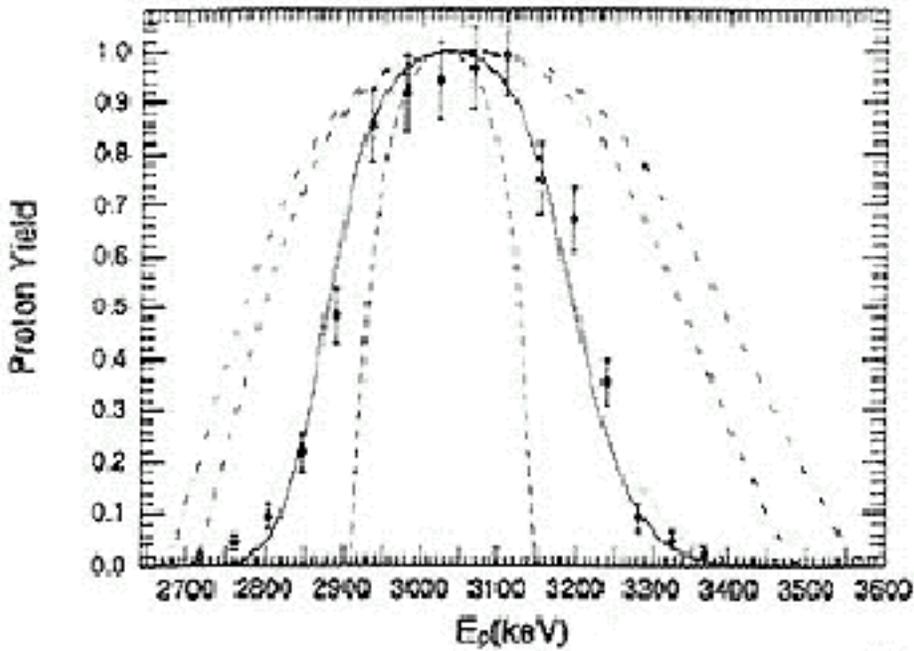

FIG. 1. Proton Energy spectrum calculated from Eq. (8) (monoenergetic D$^+$) with $E_d = 20$ keV (dashed curve), 206 keV (dot-dashed curve), and 275 keV (dot-dot-dashed curve); from Eq. (7) with $kT_e = 20$ keV (solid curve), and the case where the deuteron velocity vectors are oriented within a 25$^o$ forward cone (dotted curve). Circles with error bars are experimental data [2]. A calibration of 9 keV/channel is obtained from the energy peak of 3.025 MeV at channel 335.

For an incident deuteron with laboratory kinetic energy $E_d$, the emitted proton energy $E_p(\Theta_L)$ for the D(D,p)T reaction is given by

$$E_p(\Theta_L) = \left[a + \left(a^2 + b^2\right)^{1/2}\right]^2, \text{ where } a = \left(m_d m_p E_d\right)^{1/2} \cos\Theta_L \big/ \left(m_t + m_p\right), \text{ and}$$

$b = \left[(m_t - m_d)E_d + m_t Q\right] \big/ \left(m_t + m_p\right)$ with $Q = 4.033$ MeV. For a circular detector of radius $r_o$ at a distance $l_o$ from the target, we define the proton spectral probability function for a monoenergetic $(E_d)$ beam of $D^+$ (assuming an isotropic angular distribution for protons in the c.m. frame) as

$$P_{E_p}(E_d) = n_D' \int_0^{E_d} \frac{\sigma(E_{c.m.})}{|dE/dx|} F_S\left(E_p, E(\Theta_L), \Theta_d\right) dE, \tag{3}$$

where $F_S\left(E_p, E(\Theta_L), \Theta_d\right)$ is a proton spectral distribution function given by

$$F_S\left(E_p, E(\Theta_L), \Theta_d\right) = \frac{1}{2\pi} \cos^{-1}\left\{\frac{l_o - R_o \cos\Theta_L \cos\Theta_d}{R_o \sin\Theta_L \sin\Theta_d}\right\} \left|\frac{d\Omega_C}{d\Omega_L}\right| \left|\frac{d\cos\Theta_L}{dE_p}\right|, \tag{4}$$

where $R_o = \left(l_o^2 + r_o^2\right)^{1/2}$, $\Theta_d$ is the angle between the beam direction and the detector orientation.

$$\left|\frac{d\cos\Theta_L}{dE_p}\right| = \frac{(m_t + m_p) + \left[(m_t - m_d)E_d + m_t Q\right]/E_p}{4(m_d m_p E_d E_p)^{1/2}}, \tag{5}$$

$$\left|\frac{d\Omega_C}{d\Omega_L}\right| = \frac{(1 + \gamma^2 + 2\gamma \cos\Theta_C)^{3/2}}{|1 + \gamma \cos\Theta_C|}, \tag{6}$$

with $\cos\Theta_C = \cos\Theta_L \left(1 - \gamma^2 \sin^2\Theta_L\right)^{1/2} - \gamma \sin^2\Theta_L$

and $\gamma = \left(m_d m_p E_d\right)^{1/2}\left[m_t\left(m_t + m_p\right)Q + m_t\left(m_t + m_p - m_d\right)E_d\right]^{-1/2}$.

For the deuteron velocity distribution given by Eq. (1), the differential fusion proton spectrum is

$$Y_{kT_e}(E_p) = N_{kT_e} \frac{1}{kT_e} \int_0^\infty e^{-E/kT_e} P_{E_p}(E) dE$$

$$= N_{kT_e} n_D' \int_0^\infty \frac{S(E_{c.m.})}{E_{c.m.}|dE/dx|} e^{-\left(E/kT_e + E_G^{1/2}/E^{1/2}\right)} F_S\left(E_p, E(\Theta_L), \Theta_d\right) dE \tag{7}$$

with $N_{kT_e}$ chosen so that the maximum value of $Y_{kT_e}(E_p)$ is normalized to unity. The proton spectrum calculated with Eq. (7) and $kT_e = 20$ keV (solid curve in Fig. 1) agrees well with the experimental data [2], and implies that an ET temperature of 20 keV was achieved for the case of 275 keV $(D_2O)_{115}^+$ impacting on the $C_2D_4$ target.

For a monoenergetic deuteron beam with laboratory kinetic energy $E_d$, the differential fusion proton spectrum as a function of $E_p$ is given by

$$Y_{E_d}(E_p) = N_{E_d}P_{E_p}(E_d),\qquad(8)$$

where $N_{E_d}$ is chosen so that the maximum value of   is normalized to unity. The proton energy spectra are calculated and plotted in Fig. 1, using Eq. (8) with $E_d = 20$ keV (dashed curve), 206 keV (dot-dashed curve), and 275 keV (dot-dot-dashed curve), and Eq. (7) with $kT_e = 20$ keV  (solid curve).

As can be seen from Fig. 1, the shape and width of the proton spectrum due to monoenergetic 20 keV $D^+$ (or 200 keV $D_2O^+$ or 220 keV $D_3O^+$) contaminants (dashed curve, representing the kinematic contribution) are different from the data [2] and the result calculated from Eq. (7) for a deuteron velocity distribution, Eq. (1), with $kT_e = 20$ keV  (solid curve) and thus can be ruled out.

Using electrostatic deflection, it was shown that ions of less than about 75% (206 keV) of the full energy (275 keV) could not contribute to the fusion proton yield [2]. The remaining high-energy ( 206 - 275 keV) $D^+$ contaminants can be ruled out since the calculated proton spectra (dot-dashed and dot-dot-dashed curves in Fig. 1) for 206 - 275 keV $D^+$'s are expected to have much larger values of the FWHM (540 - 620 keV) than the observed value of $\Delta E_p \approx 320$ keV.

To assess the effect of line broadening due to straggling of deuterons as they slow down in the target, we have computed the proton energy spectrum for the case in which the deuteron velocity vectors of Eq. (1) are uniformly oriented within a forward $(0^{o} \pm 25^{o})$ cone. The calculated proton spectrum (dotted curve) is plotted in Fig. 1 for comparison with the previous results (solid curve) for the forward ($0^{o}$) deuteron direction. Both results (dotted and solid curves) are nearly identical and imply that line broadening due to deuteron straggling is expected to be small compared to ET temperature and kinematic line broadening. A one-dimensional velocity distribution within the $0^{o} \pm 25^{o}$ cone is consistent with the forward direction of the c.m. velocity of the incident deuterons.

We have used a more general spectral distribution function than that given by Eq. (4) to estimate the effect of beam spreading with a cross-sectional radius $r_b$ on the proton line broadening. For a practical value of $r_b = 0.25$ cm, we find that FWHM increases by less than 10 keV (i.e. $\Delta E_p \leq 330$ keV) compared with our previous estimate of $\Delta E_p (kT_e = 20$ keV$) \approx 320$ keV. Therefore, the effect of beam spread on the proton line broadening is much smaller than the experimental value of $\Delta E_p \approx 320$ keV. Since the detector resolution enters in quadrature, its contribution to the FWHM is negligible up to 90 keV. The detector has a 0.19-$\mu$m Al protective layer over it. This adds 1.7 keV to the linewidth. A much thicker layer would still yield negligible line broadening.

Assuming that the total proton yield for $(D_2O)_{110}^{+}$ is the same as that for $(D_2O)_{115}^{+}$ with $kT_e = 20$ keV at the same value of $E_t = 275$ keV and that $kT_e$ scales linearly with $E_t$, i.e. $kT_e = E_t / A + B$, the parameters A, B, and $f_D$ in Eq. (2) can be determined by fitting the experimental data for the total proton yield $Y(n, E_t, kT_e)$ as a function of $E_t$ using Eqs. (2) and (3). Such a fit to the $(D_2O)_{100}^{+}$ $C_2D_4$ data of Beuhler et al. [2], which is shown in Fig. 2 (a), yields $f_D = 5 \times 10^{-4}$, A = 11, and B = -5 keV. The predicted form of the linear scaling $kT_e = E_t / 11 - 5$ keV can be tested by experimental measurements of the proton energy spectra as a function of $E_t$. Using the same values of $f_D$, A, and B, theoretical proton yields for $(D_2O)_{100}^{+}$ on TiD and $ZrD_{1.65}$ are calculated as our predictions and compared with the corresponding experimental data [2] in Figs. 2 (b) and 2(c), respectively. Our predicted results agree well with the experimental data for both the TiD target and the $ZrD_{1.65}$ target.

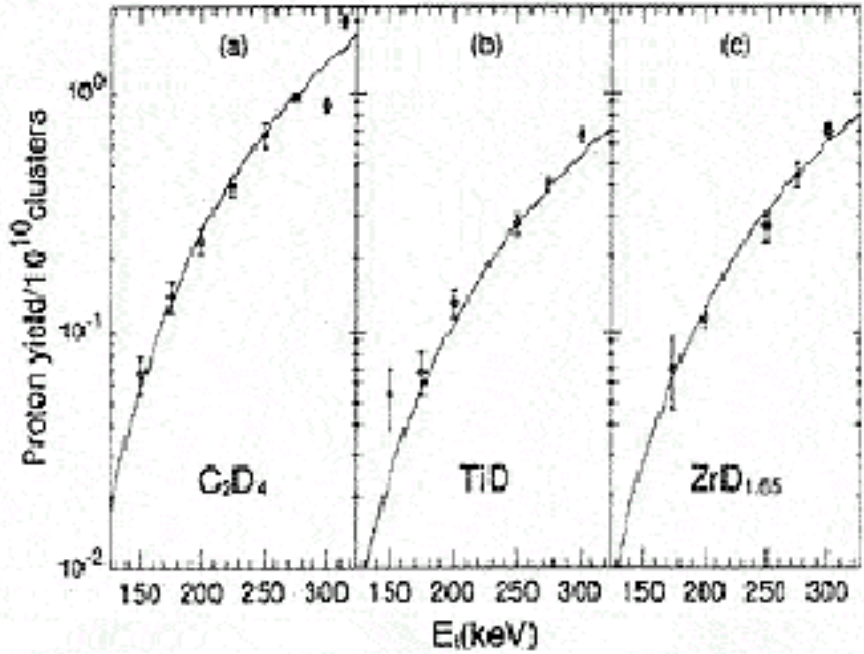

FIG. 2. Calculated fit to the $(D_2O)_{100}^+$ $C_2D_4$ proton yield data of Beuhler et al. [2] and corresponding predictions for the $(D_2O)_{100}^+$ TiD and $(D_2O)_{100}^+$ $ZrD_{1.65}$ yields as compared to the experimental data [2].

We note that $(2n)f_D$ in Eq. (2 can be less than unity since $f_D$ represents a product of two fractions, $f_D = vf_C$, where $v$ is the fraction of 2n deuterons in a cluster (i.e., $1 \le 2nv \le 2n$) and $f_C$ is the fraction of clusters in the incident beam which participate in heating deuterons to a high ET temperature $kT_e$. Energy conservation for the 1 D case requires $2nv(kT_e/2) \le E_t$. It has been observed in cluster-impact experiments [23] that energetic molecular clusters can produce craters with a diameter comparable to the size of the cluster. As stated by Beuhler et al. [2], the probability (which may be related to $f_C$) of subsequent hits of these craters by other incoming molecular clusters is not negligible. When a cluster hits the bottom of a crater, the leading edge of the projectile cluster creates upon impact a plasma consisting of target and cluster atoms (ions) and electrons whch is partially trapped in the microcrater of several Å size between the target cavity surface and the trailing cluster atoms. When the trailing cluster atoms move through this plasma, a high-velocity tail may develop for a fraction v(n) of projectile cluster and target deuterons due to mechanisms (yet to be investigated and understood) such as multiple backscattering of deuterons between target and projectile heavy atoms [8 - 10], (b) pinch instability heating due to magnetic confinement [15, 24, 25] (which favors a one-dimensional velocity distribution), (c) otgher collective effects due to electron degrees of freedom, etc.

Low energy (<20 keV) resonances are expected to yield much narrower FWHM than the data ($\Delta E_p \approx 320$ keV) for the proton energy spectrum, ruling out theoretical models based on them. Thus proton spectral broadening can be used to test theoretical models for CIF in addition to discriminating the effect of possible contaminants. Thus, it is important to measure both the total proton yield and the proton energy spectrum simultaneously in future CIF experiments.

**Acknowledgment**

We acknowledge valuable comments by Dr. Young K. Bae and Dr. Robert Vandenbosch. This work has been supported in part by the Purdue Research Foundation and EPRI.

**References**

[1] R. J. Beuhler, G. Friedlander, and L. Friedman, Phys. Rev. Lett. 63, 1292 (1989).

[2] R. J. Beuhler, Y. Y. Chu, G. Friedlander, L. Friedman, and W. Kunnman, J. Phys. Chem. **94**, 7665 (1990).

[3] Y. K. Bae, D. C. Lorents, and S. E. Young, Phys. Rev. **A 44**, 4091 (1991).

[4] M. Rabinowitz, Mod. Phys. Lett. **B4,** 665 (1990).

[5] Y. E. Kim, Fusion Technology **17**, 507 (1990).

[6] C. Carraro, B. Q. Chen, S. Schramm, and S. E. Koonin, Phys. Rev. **A 42**, 1379 (1990).

[7]  P. M. Echenique, J. R. Manson, and R. H. Ritchie, Phys. Rev. Lett. **64**, 1413 (1990).

[8]    M. Rabinowitz, Y. E. Kim, R. A. Rice, and G. S. Chulick). *AIP Conference Proceedings 228: Anomalous Nuclear Effects in Deuterium/ Solid Systems*, pp. 846-866 (1990).

[9]  Y. E. Kim, M. Rabinowitz,  G. S. Chulick and R. A. Rice,  Mod. Phys. Lett. **B5,** 427 (1991).

[10]  Y. E. Kim, R. A. Rice,G. S. Chulick,  and  M. Rabinowitz, Mod. Phys. Lett.  **A6**, 2259 (1991).

[11]  J. Fallavier, R. Kirsch, J. C. Poizat, J. Remillieux, and J. P. Thomas, Phys. Rev. Lett.   **65**, 621 (1990).

[12]  Y. E. Kim, M. Rabinowitz, Y. K. Bae, G. S. Chulick and R. A. Rice, Mod. Phys. Lett.   **B5**, 941 (1991).

[13]  R. J. Beuhler, Y. Y. Chu, G. Friedlander, L. Friedman, A. G. Alessi, V. LoDestro, and    J. P. Thomas Phys. Rev. Lett.  **67**, 473 (1991).

[14]  Y. E. Kim, G. S. Chulick, R. A. Rice, M. Rabinowitz,  and Y. K. Bae,  Chem. Phys. Lett. **184**, 465 (1991).

[15]  Y. E. Kim, M. Rabinowitz, Y. K. Bae,  G. S. Chulick and R. A. Rice, in Proceedings of      the International Symposium on Hot Plasma Shock-Wave Theory of Cluster-Impact Fusion, Nikko, Japan, 6-8 June 1991.

[16]  Y. E. Kim, Y. K. Bae, G. S. Chulick and R. A. Rice). Fusion Technology, 20 , 797-807 (1991).

[17]  M. I. Haftel, "Molecular Dynamics of 300 keV $(D_2O)_{100}$ Clusters on TiD" (unpublished).

[18]  M. Hautala, Z. Pan, and P. Sigmund, Phys. Rev. **A44**, 7428 (1991).

[19]  O. H. Crawford, "Cluster-Impact Fusion: Yields from Binary-Collision Sequences" (to be published).

[20]  H. Brysk, Plasma Phys. **15**, 611 (1973).

[21]  H. H. Anderson and J. F. Ziegler, *Hydrogen Stopping Powers and Ranges in All Elements* (Pergamon, New York, 1977).

[22]  W. A. Fowler, G. R. Caughlan, and B.A. Zimmermann, Annu. Rev. Astron. Astrophys. **5**, 5 25 (1967); **13**, 69 (1975).

[23]  M. W. Matthew,  R. J. Beuhler, M. Ledbetter, and L. Friedman, Nucl. Instrum. Methods Phys. Res. Sect. B **14**, 448 (1986).

[24]  A. Hasegawa et al., Phys. Rev. Lett.  **56**, 139 (1986).

[25]  Y. E. Kim, in *Laser Interaction and Related Phenomena,*  International Conference   Series edited by  H. Hora and G. H. Miley (Plenum, New York, 1989), pp. 583-592.